\documentclass[smallextended]{svjour3}       
\smartqed  
\usepackage{amsmath,amssymb}
\usepackage{graphicx}
\usepackage{xcolor}
\newcommand{\ket}[1]{\vert #1 \rangle}
\newcommand{\bra}[1]{\langle #1 \vert}

\newcommand{\braket}[2]{\langle #1 \vert #2 \rangle}

\newcommand{\ketbra}[2]{\vert #1 \rangle  \langle #2 \vert}

\journalname{Quantum Information Processing}
\begin{document}
\title{Process estimation in qubit systems: a quantum decision theory approach}
\subtitle{}
\titlerunning{Process estimation in qubit systems}        

\author{Ivan Maffeis \and Seid Koudia \and Abdelhakim Gharbi \and 
Matteo G. A. Paris}
\authorrunning{I. Maffeis {\em et al.}} 

\institute{I. Maffeis and M. G. A. Paris \at
Quantum Technology Lab, Dipartimento di Fisica 
{\em Aldo Pontremoli}, Universit\`a degli Studi di Milano, I-20133 
Milano, Italy \and 
S. Koudia and A. Gharbi \at
Laboratoire de Physique Th\'eorique, Facult\'e des Sciences Exacte,
Universit\'e de Bejaia, 06000 Bejaia, Algeria}
\date{\today}
\maketitle
\begin{abstract}
We address quantum decision theory as a convenient framework to 
analyze process discrimination and estimation in qubit systems. 
In particular we discuss the following problems: i) how to 
discriminate whether or not a given {\it unitary perturbation}
has been applied to a qubit system; ii) how to determine the 
amplitude of the {\it minimum detectable perturbation}. In order 
to solve the  first problem, we exploit the so-called Bayes 
strategy, and look for the optimal measurement to discriminate, 
with minimum error probability, whether or not the unitary 
transformation has been applied to a given signal. Concerning 
the second problem, the strategy of Neyman and Pearson is used 
to determine the ultimate bound posed by quantum mechanics to 
the minimum detectable amplitude of the qubit transformation. 
We consider both pure 
and mixed initial preparations of the qubit, and solve the 
corresponding binary decision problems. We also analyze the use 
of entangled qubits in the estimation protocol and found that 
entanglement, in general, improves stability rather
than precision. Finally, we take into account the possible
occurrence of different kinds of background noise 
and evaluate the corresponding effects 
on the discrimination strategies. 
\end{abstract}
\section{Introduction}\label{intro}
The existence of non-orthogonal quantum states is one of the 
fundamental traits of quantum mechanics, and has profound 
implications on its applications. On the one hand, non-orthogonality
poses limitations to the fundamental challenge of quantum state discrimination 
\cite{yue75,hel67,hel68,hel76,iva87,die88,per88,bgu04,che04,tom08,bgu10,jex03,jex04,jex06} and, on the other hand, it may be exploited as a resource in quantum technologies, e.g. in quantum cryptography \cite{ben92,eke91,gol95,guo95,par00,luc05,ben14,sca14,ave10,bri12,mes0,mes1,mes2}. 
Non-orthogonality also influences 
other fundamental tasks in quantum information processing, and in 
particular process discrimination \cite{dar02,lai09,inv10,dev12,wit14,bae15,tra17,reh18,che18}, 
which itself represents a crucial ingredient for quantum 
simulations and quantum interferometry \cite{tak10,int97,ral00}.
\par
In this paper, we address process discrimination in qubit systems 
and exploit results from quantum decision theory in order to 
optimize the discrimination strategy \cite{dua08,cao16}, 
i.e. to minimize the impact of non-orthogonality, and 
to derive the corresponding ultimate bounds to the
{\it probability of error} in the detection of the 
perturbation, and the {\it discrimination precision} of the
perturbation amplitude. The possible advantages resulting from the 
use of entanglement are also  explored in details.  
\par
The scheme we are going to consider is the following: a single- or 
two-qubit system is prepared in a given initial state, then, 
with a certain unknown probability, it is subjected to the action of 
a given unitary operator $U_\lambda = \exp \{ - i G \lambda\}$, 
$G$ being the {\em generator} of the perturbation and $\lambda$ 
its {\em amplitude}. Finally, the system is measured, in order to 
detect whether or not the unitary transformation has perturbed 
the system. The problem is equivalent to that of discriminating 
the unperturbed state of the system from the perturbed one, while 
accepting a probability of error. 
A second, complementary, goal is to determine the minimum detectable 
value of the amplitude $\lambda$ which leads to discriminable outputs. 
The precision obtained by using one- or two-qubit will 
be compared in order to reveal whether the use of entanglement 
leads to some advantages, either in ideal conditions or 
in the presence of noise.
\par
The paper is structured as follows. In Section \ref{sec2} we establish
notation and briefly review few fundamental results in quantum 
discrimination theory, also illustrating the different figures 
of merit employed in the so-called {\em Bayes} and {\em Neyman-Pearson} 
discrimination strategies. In Section \ref{sec3} we employ Bayes strategy 
to distinguish between different processes with minimum error using 
one- and two-qubit probes, also when different kind of noise occurs.
In Section \ref{sec4}, we exploit Neyman-Pearson strategy to find the 
ultimate bound to the minimum detectable perturbation, discrimination, 
also addressing the use of entanglement and the occurrence of noise.
Finally, Section \ref{outro} is closing the paper with some concluding remarks.    
\section{Quantum decision theory\label{sec2}}
Let us consider a qubit system and assume it may be prepared 
in one of the two states $\left\{ \rho_{j}\right\} _{j=1,2}$ with 
prior probabilities $\left\{ z_{j}\right\} _{j=1,2}$ 
\begin{color}{black}
such that 
$\rho_{j}\in\mathcal{L} \left({\mathbb C}^2\right)$, and 
$\sum_{j}z_{j}=1$. 
\end{color}
Our goal is that of inferring the state of the qubit, basing 
our decision on the outcome of a measurement performed 
on the system. To this aim we should implement a detection 
scheme,
i.e. a POVM (Positive Operator-Valued Measure) 
$\Pi=\left\{ \Pi_{k}\right\} _{k=1,2}$, where the $\Pi_{k}$'s are 
positive semi-definite, $\hbox{Tr}\left[ \rho\,\Pi_{k} \right]\geq0$, $\forall \rho$, and sum to the identity $\sum_{k}\Pi_{k}=\mathbb{I}$. 
\par
As we already mentioned, we cannot, even in principle, perfectly distinguish the two states, unless they have orthogonal supports. For 
non-orthogonal states, we cannot achieve perfect discrimination. However, we may seek for an {\em optimal discrimination strategy}, 
which minimise/maximise, in average, a given loss/gain function. 
In the following
two paragraphs, we briefly review two relevant strategies 
used in quantum hypothesis testing, which will be also at the basis 
of our approach to process estimation in qubit systems.
\par
The first paragraph is devoted to Bayes strategy, which aims to
the POVM minimising the average probability of error in the decision process. Bayes strategy is suitable for those situations where the two possible outcomes of the measurement are equally important for the experimenter and thus the
two error probabilities should be jointly minimised. The paradigmatic situation where the Bayes strategy is employed is binary quantum communication, in which two classical symbols
are encoded onto quantum states of a physical system and 
quantum decision theory is exploited to determine the optimal 
receiver at the output of the communication channel.
\par
On the other hand, there are several situations of interest where
one of the two events, usually referred to the {\em alternative hypothesis}, is expected to occur more rarely with respect to the other one, called {\em null hypothesis}.
In these cases, the detection of alternative hypothesis is the main task of the measurement.
The so-called Neyman-Pearson (NP) strategy is relevant for this context, aiming at maximising the detection 
probability of the alternative hypothesis while accepting a possible {\em false-alarm probability}, which is the probability of inferring the alternative hypothesis when the  null hypothesis is instead true.
A possible paradigmatic situation where NP strategy may be successfully implemented is the interferometric detection of gravitational waves.
\subsection{Bayes discrimination strategy}
Bayes strategy aims at minimising the average probability 
of error of the decision process \cite{bgu10}, i.e. 
the quantity $$q_{e}(\Pi)=\sum_{i\neq j}z_{j}\,p\left(i|j\right)\,,$$ 
where $p\left(i| j\right)=\mathrm{Tr}\left[\rho_{j}\Pi_{i}\right]$ is
the probability of inferring $\rho_{i}$ when $\rho_{j}$ is the actual state of the system. In turn, Bayes strategy is also referred to as {\em minimum-error} discrimination strategy \cite{mes0,mes1,mes2}.
Since we are here considering 
binary decisions, the probability of error simplifies to 
\begin{align}
  q_{e} (\Pi)& = z_{1}- \hbox{Tr}\left[ \Lambda\,\Pi_{1} \right]
  = z_0 + \hbox{Tr}\left[ \Lambda\,\Pi_{0} \right],
\label{eq: Prob err}
\end{align}
where 
\begin{equation}
  \Lambda=z_{1}\rho_{1}-z_{0}\rho_{0}.
  \label{eq: Charachteristic Op}
\end{equation}
This operator is hermitian, though not positive definite. It 
is sometimes referred to as the {\em Bayesian characteristic 
operator}.
\par
As we can see from (\ref{eq: Prob err}), the probability of error
is minimized when the measurement operator $\Pi_{1}$ is the 
projector on the positive part of $\Lambda$. It is thus a 
projective valued measure (PVM) given 
by $\Pi=\left\{ \mathbb{I}-\Pi_{1},\Pi_{1}\right\}$ where $\Pi_{1}=\sum_{\lambda_{k}>0}\ketbra{\lambda_{k}}{\lambda_{k}}$ and $\left\{\lambda_{k}\right\}_{k=0,1}$ are
the eigenvalues of $\Lambda$. The resulting minimum 
probability of error may be written as:
\begin{equation}
p_{e}\equiv \min_\Pi q_e(\Pi) = \frac12 \left(1 - 
\hbox{Tr} \left|\Lambda\right| \right)
= \frac12 \left(1- \lVert z_{1}\rho_{1}-z_{0}\rho_{0} \rVert_1\right)\,,
\label{eq: HE}
\end{equation}
where the trace norm $\lVert A \rVert_1$ of an operator is defined as
$\lVert A \rVert_1 = \hbox{Tr} |A| = \hbox{Tr}\left[\sqrt{A^\dag A}\right]$.
The quantity $p_e$ in Eq.~(\ref{eq: HE}) is usually referred to as 
the {\em Helstrom bound} to the
error probability in binary state discrimination.
For pure states the Helstrom bound may be rewritten
as
\begin{equation}
p_{e}=\frac{1}{2}\left[1-\sqrt{1-4z_{0}z_{1}\left|\kappa
\right|^{2}}\right], \label{eq:prober}
\end{equation}
where $\kappa=\left\langle \psi_{1}|\psi_{0}
\right\rangle$ is the overlap between the two states.
We can easily see that for vanishing overlap the probability of
error is also vanishing as $p_e \stackrel{|\kappa|\rightarrow 0}{\simeq} z_{0}z_{1}\left|\kappa
\right|^{2}$, whereas for large $\kappa$ ($\kappa\rightarrow 1$), the discrimination process approaches pure guessing ($p_e \rightarrow 1/2$).
\subsection{Neyman-Pearson discrimination strategy}
Neyman-Pearson strategy aims at maximising the detection 
probability of the alternative hypothesis, at a fixed value of the 
false-alarm probability. Let us refer to the 
two hypotheses $\left\{ H_{0},H_{1}\right\} $ as the
null and the alternative hypothesis respectively, where 
$H_{j,\, j=0,1}$ corresponds to the qubit being in the state $\rho_{j}$. 
The basis of NP strategy is in fact a trade-off between the probability of detection of alternative hypothesis, $p(1|1)\equiv p_{11}$, and the false-alarm probability, $p(1|0)\equiv p_{10}$. A maximum threshold of false-alarm probability is fixed as acceptable, given the nature of the physical problem where the decision strategy is implemented.
The searching of the POVM maximising the
detection probability $p_{11}$ corresponds to a Lagrange maximisation problem where the value of $p_{10}$ is taken as
a constraint. The optimal POVM is then the one maximising the Lagrange
functional \cite{hel76}:
\begin{align}
L =p_{11}-\gamma p_{10} 
  =\hbox{Tr}\left[ \Gamma\,\Pi_{1} \right],\nonumber
  \label{eq:lagrange problem}
\end{align}   
where $\gamma$ is a Lagrange multiplier and 
\begin{equation}
\Gamma=\rho_{1}-\gamma\rho_{0}\label{eq:lagrange operator}
\end{equation}
is the Lagrange operator. In order to maximise $L$, the POVM $\Pi_{1}$
needs to be a projector on the positive eigenvalues of the operator
$\Gamma$. The optimal measurement scheme according to 
Neyman-Pearson strategy is thus a projective one 
$\Pi=\left\{ \mathbb{I}-\Pi_{1},\Pi_{1}\right\}$ where 
$\Pi_{1}=\sum_{g>0} |g\rangle\langle g|$, $g$ being 
the eigenvalues of $\Gamma$ and $|g\rangle$ the corresponding eigenvectors.
We notice that, in solving the
eigenvalue problem, different values of $\gamma$ correspond to different
values of the accepted false alarm probability and thus to different Neyman-Pearson strategies.
\par
Once the eigenvalues of $\Gamma$ are found, the strategy becomes
clear: when the measurement of $\Gamma$ gets a positive outcome, the alternative hypothesis $H_{1}$ is inferred as true, otherwise the null hypothesis $H_{0}$ is inferred. For pure states the detection
probability $p_{11}$ can be written in terms of $p_{10}$ after eliminating the
Lagrange multiplier $\gamma$, obtaining: 
\begin{equation}
p_{11}\left(p_{10}\right)=
\left\{
\begin{array}{cl}
\biggl[\sqrt{p_{10}\left|\kappa\right|^{2}}+\sqrt{\left(1-p_{10}\right)\left(1-\left|\kappa\right|^{2}\right)}\biggr]^{2} & 0\leq p_{10}\leq\left|\kappa\right|^{2},\\
 & \\
1 & \left|\kappa\right|^{2}<p_{10}\leq 1.
\end{array}\right.\label{eq:NP pure}
\end{equation}
Eq.~(\ref{eq:NP pure}) shows that we may have unit detection probability
as far as we accept a false-alarm probability larger than the
overlap between the two pure states involved in the problem. 
As a consequence, as the overlap is large, the detection of 
the alternative hypothesis becomes hard, as either the detection probability cannot achieve the unit value without accepting extremely large margins of error (false alarm).
\section{Bayes approach to process detection in qubit systems\label{sec3}}
\subsection{Single-qubit states}
\begin{figure}[h!]
\includegraphics[width=0.95\columnwidth]{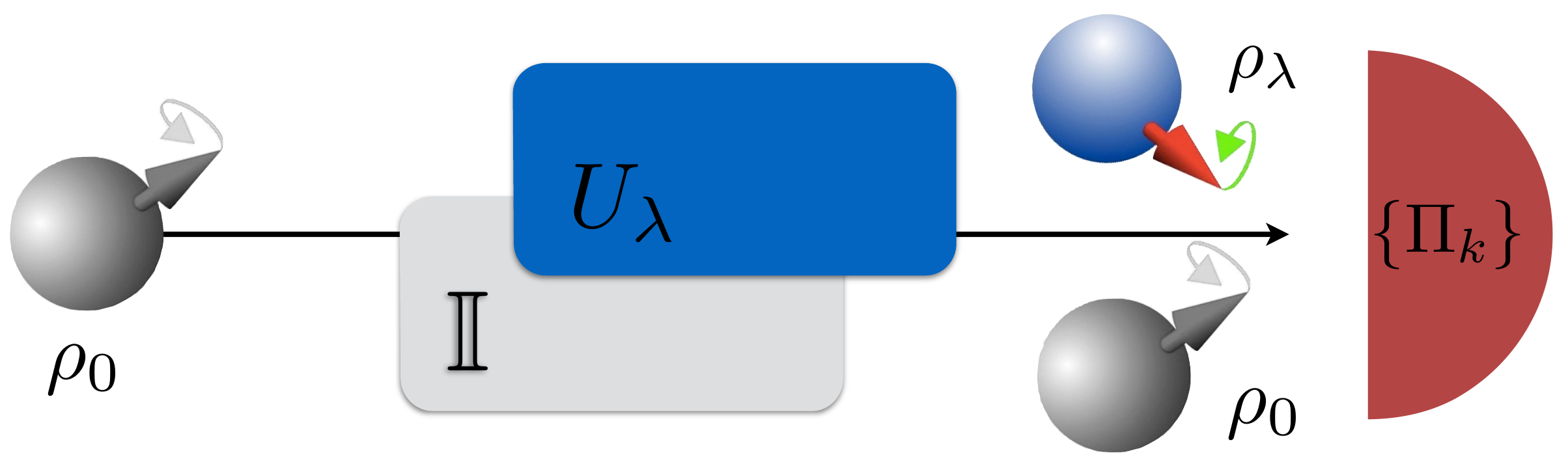}\\
\caption{Process discrimination as a binary decision
problem. A single qubit initially prepared in the quantum state 
$\rho_{0}$ may, or may not, undergo a unitary transformation $U_{\lambda}$. A detector is placed at the output of the 
system to discriminate between the two states and, in turn, 
whether the unitary transformation has occurred or not.\label{fig:scheme1q}}
\end{figure}

In this section we apply Bayes strategy to quantum binary discrimination 
to observe whether or not a given unitary perturbation has been applied
to a qubit system. The detection scheme is schematically illustrated 
in \textbf{Fig.~\ref{fig:scheme1q}}.
\par
A single-qubit system is initially prepared in an 
input quantum state $\rho_{0}$ and it is then let 
to evolve through a channel where it may 
undergo a transformation $U_{\lambda}$, being 
$\lambda$ the perturbation amplitude. At the 
output, i.e .after the (possible) transformation 
we want to know whether we have $\rho_{0}$ or 
\begin{equation}
\rho_{\lambda}=U_{\lambda}^{\dagger}\rho_{0}U_{\lambda}
\label{eq:perturbed state}
\end{equation}
as output state.
The problem reduces to discrimination between the two states, 
and since they do not have, in general, 
orthogonal supports, quantum decision theory is 
the natural framework to adopt. In the following, 
we will employ Bayes strategy for single-qubit signals, 
as well as two-qubit ones, in order to detect perturbations
induced by a Pauli matrix and, without loss of generality, 
we will assume
\begin{align}
U_{\lambda} =e^{-i\lambda\sigma_1}\label{eq:generic transfor}
=\cos\lambda\,\mathbb{I}-i\,\sin\lambda\,\sigma_1
\end{align}
as the transformation which may take place into the channel.
Upon writing the initial single-qubit state in a Bloch form  
$\rho_{0}=\frac{1}{2}\left(\mathbb{I}+\mathbf{r} \cdot
{\boldsymbol \sigma}\right)$,
with $\mathbf{r} = \{r_1,r_2,r_3\}$, ${\boldsymbol \sigma} = \{\sigma_1, \sigma_2, \sigma_3\}$, the transformed state is given by 
$\rho_{\lambda}=\frac{1}{2}\left(\mathbb{I}+\mathbf{r}_\lambda \cdot
{\boldsymbol \sigma}\right)$
where the transformed components are given by 
\begin{align}
r_{1\lambda} & =r_{1},\nonumber \\
r_{2\lambda} & =r_{2}\cos 2\lambda - r_{3}\sin 2\lambda, \\
r_{3\lambda} & =r_{2}\sin2\lambda +  r_{3}\cos 2\lambda\nonumber.\label{eq:components}
\end{align}
The characteristic operator in (\ref{eq: Charachteristic Op}) becomes
$\Lambda=\begin{pmatrix}\Lambda_{0} & \Lambda_{2}\\
\Lambda_{2}^{*} & \Lambda_{1}
\end{pmatrix}
$,
with:
\begin{alignat}{1}
\Lambda_{0} & =z_{1}\left(1+r_{3\lambda}\right)-z_{0}\left(1+r_{3}\right),\nonumber \\
\Lambda_{1} & =z_{1}\left(1-r_{3\lambda}\right)-z_{0}\left(1-r_{3}\right),\\
\Lambda_{2} & =z_{1}\left(r_{1\lambda}-i\,r_{2\lambda}\right)-z_{0}\left(r_{1}-i\,r_{2}\right).\nonumber 
\end{alignat}
In this context, $z_1$ and $z_0$ are respectively the prior probabilities 
for the unitary transformation to have effect on the initial quantum state 
or not ($z_0 + z_1 = 1$).
Upon evaluating explicitly trace and determinant of $\Lambda$, the 
probability of error has the following form:
\begin{equation}
p_{e}=\frac{1}{2}\left[1-\sqrt{r^{2}-4z_{1}z_{0}\left[r^{2}-\left(r^{2}-r_{1}^{2}\right)\mathrm{sin^{2}\lambda}\right]}\right],\label{eq:p error}
\end{equation}
\textcolor{black}{in which $r^{2}={\left|\mathbf{r}\right|}^{2} = r_1^2+r_2^2+r_3^2$.}
If the state is pure,  $r^{2}=1$, we indeed recover 
Eq. (\ref{eq:prober}) since we have $\left|\kappa\right|^{2}
=1-\left(1-r_{1}^{2}\right) \mathrm{sin^{2}\lambda}$. 
In the limiting cases where both eigenvalues are positive (negative)
discrimination reduces to pure guessing, i.e. the inference is made 
without looking at data. We may summarize the situation as follows: 
\begin{gather}
\hbox{Tr}\left[\Lambda\right]\leq0\mathrm{\hfill\qquad Tr\left[\Lambda\right]>0}\nonumber \\
\Pi=\left\{ \Pi_{0}=\mathbb{I},\Pi_{1}=\mathbb{O}\right\} \hfill\qquad\Pi=\left\{ \Pi_{0}=\mathbb{O},\Pi_{1}=\mathbb{I}\right\} \label{eq:povm}\\
p_{e}=z_{1}\,\hfill\qquad p_{e}=z_{0}\nonumber 
\end{gather}
\par
Let us now assume $z_{1}=z_{0}=\frac{1}{2}$ (no {\em a priori} 
information about the occurrence of the perturbation) and rewrite 
(\ref{eq:p error}) in terms of the initial state purity
\begin{equation}
p_{e}=\frac{1}{2}\left[1-\sqrt{\left(2\mu-1- r_1^{2}\right)\mathrm{sin}^{2}\lambda}\right]\label{eq:p erro purity+projection}\,.
\end{equation}
The probability of error achieves its minimum when the input signal
is pure ($\mu=1$), and its projection on the direction of the generator
vanishes, i.e. for pure states lying in the orthogonal plane of 
the generator $\sigma_{1}$. In this case we have 
\begin{equation}
p_{e}=\frac{1}{2}\left(1-\left|\mathrm{sin\lambda}\right|\right)\label{eq: minimum error}
\end{equation}
which only depends on the parameter of the transformation. This is
the least possible error we can get. Any discrimination procedure in the Bayesian sense is thus preparation dependent, at least for the present
single-qubit case. We also note that all the above results generalise,
{\em mutatis mutandis}, to any generator $\sigma_k$ ($k=1,2,3$).
\subsection{Two-qubit states}
\begin{figure}[h!]
\includegraphics[width=0.95\columnwidth]{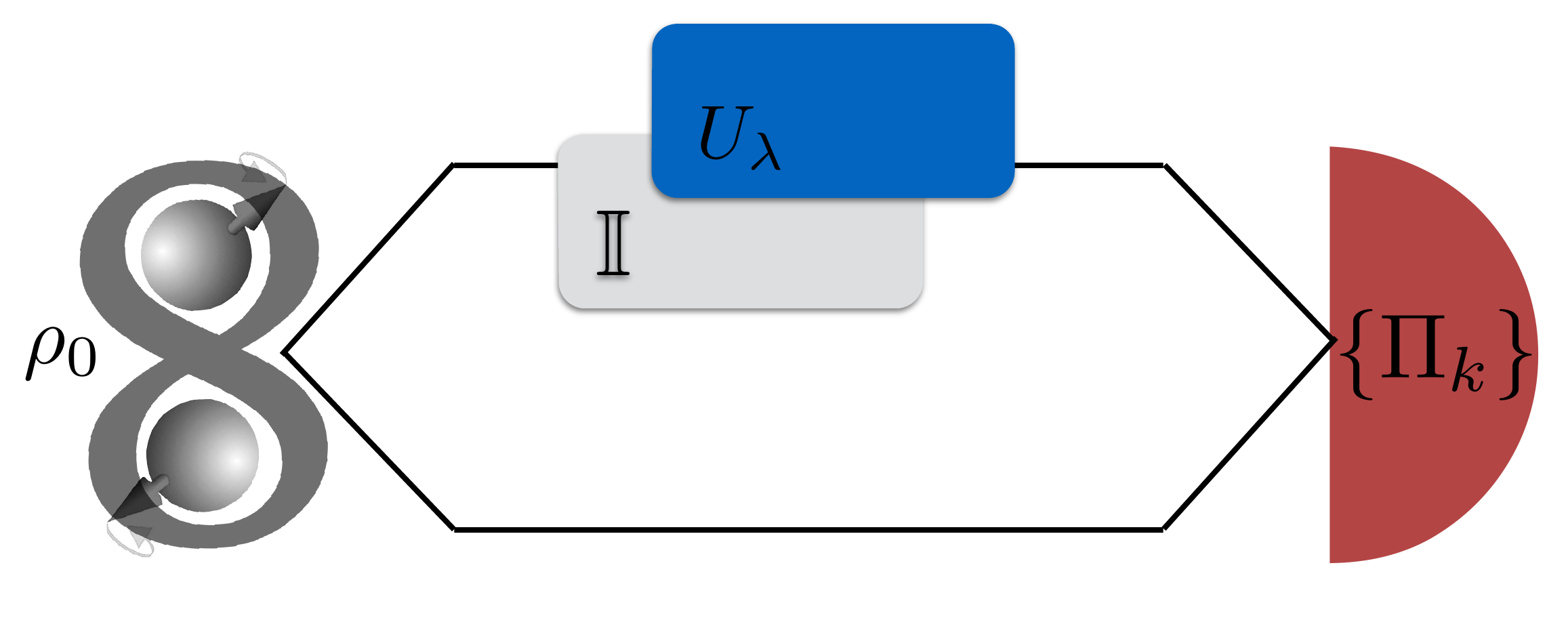} 	
\caption{Process discrimination as a binary decision
	on entangled state: the transformation $U_{\lambda}$
	that may or may not perturb the system,
	is acting on a qubit which is part of a bipartite 
	system prepared in an entangled state. At the output, a joint 
	detector is placed.\label{fig:The-experimental-protocol}}
\end{figure}

Let us now consider the scheme in 
\textbf{Fig.~\ref{fig:The-experimental-protocol}}, where 
we assume that the qubit which may possibly subject to 
the transformation $U_{\lambda}$ is initially prepared 
in an entangled state with another qubit. 
\par
Our goal is to compare the performance of this scheme against those 
if the single-qubit one 
at fixed use of the resource, here intended as the device that 
may impose the perturbation. 
Starting with a singlet state \textcolor{black}{$\ket{\psi_{-}}=\left(\ket{01}-\ket{10}\right)/\sqrt{2} $}, the density 
matrix of the corresponding  perturbed state may be written as 
\textcolor{black}{\begin{align}
	\rho_{\lambda} =\, & \cos^2\lambda\, \ketbra{\psi_-}{
		\psi_-} + \sin^2\lambda\, \ketbra{\phi_+}{\phi_+} 
	\nonumber \\
	& + i\,\sin\lambda \cos\lambda\, \left(\ketbra{\psi_-}{\phi_-} - \ketbra{\phi_-}{\psi_-}\right)
	\label{eq:perturbed state-1}\, ,
	\end{align}
where the perturbation $U_{\lambda}$ is acting on 
the first qubit in the singlet.} The Bayesian characteristic 
operator is thus given by
\textcolor{black}{\begin{align}
	\Lambda  = & \label{eq:Bayesian characteristic}
	\left(z_1\cos^{2}\lambda-z_0\right) \ketbra{\psi_-}{\psi_-}+
	\sin^2\lambda \, \ketbra{\phi_{+}}{\phi_{+}} \\  
	& + i\,\sin\lambda\, \cos\lambda\ \left(\ketbra{\psi_-}{\phi_-} - \ketbra{\phi_-}{\psi_-}\right) \nonumber\,,
	\end{align}}\\ 
and the corresponding eigenvalues by 
$\xi_{\pm}=\frac{1}{2}\Big[\left(z_{1}-z_{0}\right)\pm\sqrt{1-4z_{0}z_{1}\cos^{2}\lambda}\Big]$.
The error probability reads as follows:
\begin{equation}
p_{e}=z_{1}-\xi_{+}=\frac{1}{2}\Bigg[1-\sqrt{1-4 z_{0}z_{1}\cos^{2}\lambda}\Bigg]\label{eq:prob err}\,.
\end{equation}
\textcolor{black}{
Using entanglement we may thus achieve the ultimate bound
in Eq. (\ref{eq: minimum error}) and, remarkably,
the same results, i.e. the same value of $p_{e}$ in
Eq.~(\ref{eq:prob err}) is obtained for any initial 
two-qubit preparation of the form
\begin{align}
	\ket{\psi}=\frac{1}{\sqrt{2}}\left[\alpha_0\ket{00}+\alpha_1\ket{01}+\alpha_2\ket{10}+\alpha_3\ket{11}\right]\label{eq:general 2-qubit}\,,
	\end{align}
provided that $\alpha_0\alpha_2^{*}+\alpha_1\alpha_{3}^{*}=0$.
In addition, under the same condition, 
Eq.~(\ref{eq:prob err}) also holds when the perturbation
is generated by any of the $\sigma_k$, $k=1,2,3$.
The class of states in Eq. (\ref{eq:general 2-qubit})
includes maximally entangled states. 
Upon comparing Eq. (\ref{eq:general 2-qubit}) with
Eq.~(\ref{eq:p erro purity+projection}), we see that entanglement 
improves the discrimination process in terms of {\it stability}, 
since the probability of error does not depend on the projection 
of the Bloch vector on the axis corresponding to the generator 
of the perturbation.} 
\par
Let us now consider a generic mixture of Bell states:
\textcolor{black}{\begin{equation}
	\rho_{0}=p_{0}\ket{\phi_{+}}\bra{\phi_{+}}+p_{1}\ket{\psi_{+}}\bra{\psi_{+}}+p_{2}\ket{\psi_{-}}\bra{\psi_{-}}+p_{3}\ket{\phi_{-}}\bra{\phi_{-}}\,. \nonumber 
	\end{equation}}\\ 
After lengthy but straightforward calculation the characteristic operator can be obtained (not shown here) and, in turn, the following probability of error:
\begin{equation}
p_{e}=\frac{1}{2}\left[1-\left(\left|p_{0}-p_{1}\right|+\left|p_{1}-p_{3}\right|\right)\left|\mathrm{sin\lambda}\right|\right]\label{eq:prob error diag}\,.
\end{equation}
\subsection{Bayes strategy in presence of noise \label{subsecnoisy}}
\begin{figure}[h!]
\includegraphics[width=0.95\columnwidth]{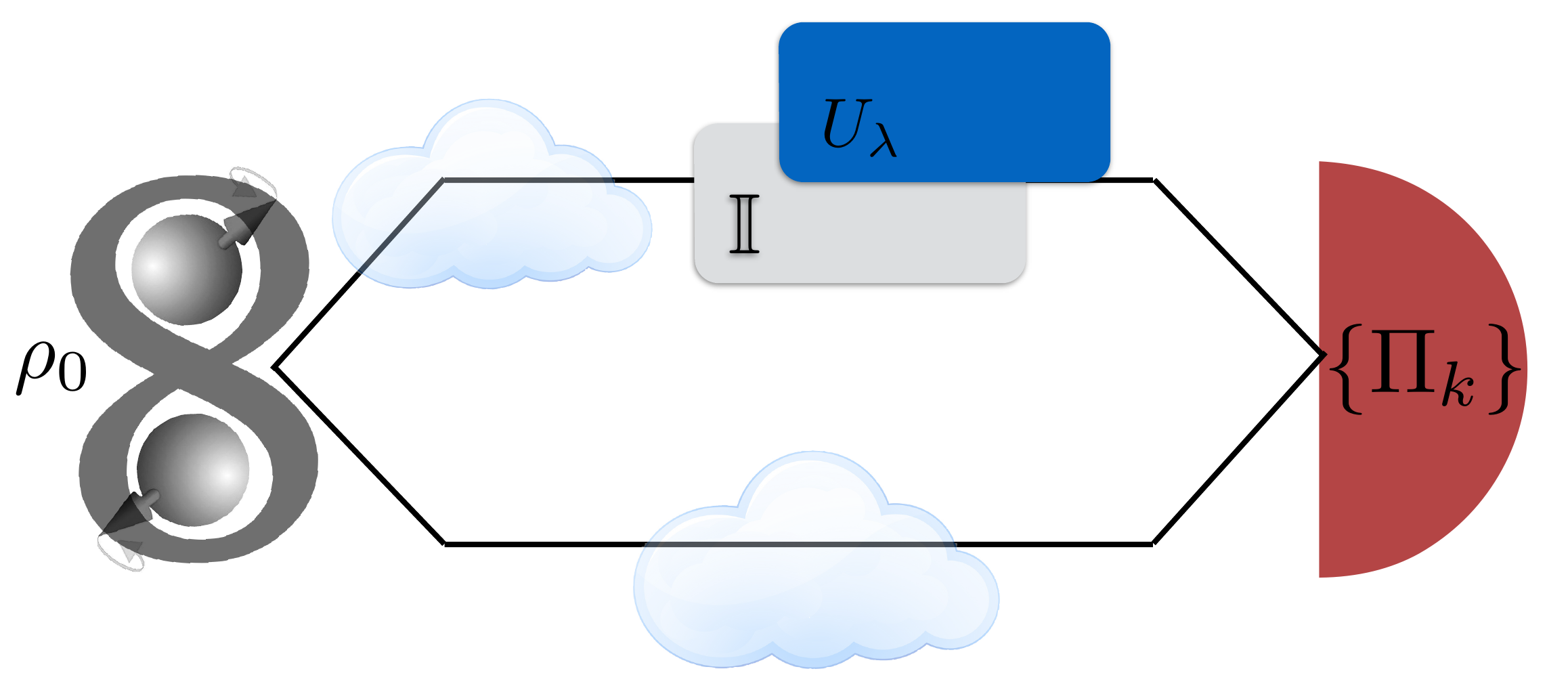} 
\caption{
Process discrimination as a binary decision on noisy 
entangled state. The signal is subject to some source 
of noise, which affects both parts of the entangled system, and takes place before the possible perturbation $U_{\lambda}$. At the output we have a joint detector.\label{fig:The-experimental-protocol-1}}
\end{figure}

A possible generalization of the problem is considering a situation where some source of noise acts during propagation of the signal.
In particular, we focus on the two-qubit case, assuming that noise occurs to both parties and takes place before the possible perturbation $U_{\lambda}$.
A schematic diagram of the experimental protocol is shown in 
\textbf{Fig.~\ref{fig:The-experimental-protocol-1}}. 
We present here the analysis of the effects on the same initial entangled state of four different kind of noise: the three formalized by Pauli matrices and the depolarizing noise.

\par
The analysis here presented considers as first case the \textit{bit-flip} noise, which is described by the completely positive map 
\begin{equation}
{\cal E}_1 (\rho_{0})=\sum_{i,j=0}^{1}E_{ij}\,\rho_{0}\,E_{ij}^{\dagger}, \label{eq:action of noise}
\end{equation}
where $E_{ij}$ are the Kraus operators:
\begin{align}
E_{00}&=\sqrt{pq}\,\mathbb{I}\otimes\mathbb{I},\nonumber \\
E_{01}&=\sqrt{p\left(1-q\right)}\,\mathbb{I}\otimes\sigma_{1},\nonumber \\
E_{10}&=\sqrt{\left(1-p\right)q}\,\sigma_{1}\otimes\mathbb{I},\nonumber \\
E_{11}&=\sqrt{\left(1-p\right)\left(1-q\right)}\,\sigma_{1}\otimes\sigma_{1}.\label{eq:Kraus Op}
\end{align}
In these equations, $p$ and $q$ are the probabilities for respectively the first qubit (which has probability to undergo the unitary transformation) and the second qubit to be subject to noise. The probability of occurring noise is assumed as independent on the two channels.
If the initial state is \textcolor{black}{$\rho_{0}=\ket{\phi_{+}}\bra{\phi_{+}}$} , we get  
\textcolor{black}{\begin{align}
	{\cal E}_1 (\rho_{0})& =  \left[pq+\left(1-p\right)\left(1-q\right)\right]\left|\phi_{+}\rangle\langle\phi_{+}\right|+\left[p\left(1-q\right)+q\left(1-p\right)\right]\left|\psi_{+}\rangle\langle\psi_{+}\right|\nonumber \\ &
	= \left(1-p-q+2pq\right)\left|\phi_{+}\rangle\langle\phi_{+}\right|+\left(p+q-2pq\right)\left|\psi_{+}\rangle\langle\psi_{+}\right|\label{eq:action of bit flip}\,.
	\end{align}} 
Using Eq.~(\ref{eq:prob error diag}) we obtain: 
\begin{equation}
p_{e}=\frac{1}{2}\left[1-\left|\left(2p-1\right)\left(2q-1\right)\right|\left|\mathrm{sin\lambda}\right|\right]\,.
\end{equation}

Similar considerations can be conducted for the \textit{phase-flip} noise by replacing $\sigma_{1}$ in Eq.~(\ref{eq:Kraus Op}) to $\sigma_{3}$.
For the same initial state considered above we get:
\begin{align}
{\cal E}_3 (\rho_{0}) & =\left[pq+\left(1-p\right)\left(1-q\right)\right]\left|\phi_{+}\rangle\langle\phi_{+}\right| +\left[q\left(1-p\right)+p\left(1-a\right)\right]\left|\phi_{-}\rangle\langle\phi_{-}\right|,\label{eq:action of phase flip} 
\end{align} 
and
\begin{equation}
p_{e}=\frac{1}{2}\left[1-\left|\mathrm{sin\lambda}\right|\right]\label{eq:prob min}\,.
\end{equation}
No dependence on the noise is present, i.e. the discrimination power for the studied protocol is not affected by the presence of phase-flip noise.
Analogue results are obtained when discussing the 
{\em phase-bit-flip noise}, which is described by the same previous set of Kraus operators by formally replacing $\sigma_{1}$ with $\sigma_{2}$.

\par
Considering as a last case the depolarizing noise 
\begin{equation}
{\cal E}_{dp} (\rho_{0})=\frac{p}{4}\mathbb{I}+\left(1-p\right)\rho_0,\label{eq:action of depol noise}
\end{equation}
where $p$ is the probability that the state $\rho_{0}$ is transformed
to a maximally mixed state $\frac{1}{2}\mathbb{I}$, results of
Eq.~(\ref{eq:prob error diag}) may be used, leading to 
\begin{equation}
p_{e}=\frac{1}{2}\left[1-\left(1-p\right)\left|\mathrm{sin\lambda}\right|\right].
\end{equation}
As before, this is independent of the mean value of the generator of the perturbation.
On the other hand, the probability of error depends on the noise parameter $p$, and it increases with $p$.
To summarize:  when the noise acts on an eigenspace of 
the perturbation generator, the decision process
is affected, and accordingly, the probability of error is increased.
On the other hand, when the noise acts on an orthogonal space 
of the generator of the perturbation $U_{\lambda}$,
the probability of error of making a decision results unchanged 
and it achieves its minimal value given in Eq.~(\ref{eq:prob min}).
\section{Neyman-Pearson strategy and the minimum detectable perturbation}\label{sec4}
\subsection{Pure states}
In this Section we address the problem of evaluating 
the minimum detectable value of the perturbation amplitude 
$\lambda_m$, i.e. the minimum value of $\lambda$ in $U (\lambda)=e^{-i \lambda\sigma_1}$ making $\rho_0$ and $\rho_\lambda$ discriminable. 
To this aim, we have to maximise the detection probability 
of the alternative hypothesis, which is the basis of the so-called Neyman-Pearson strategy, to optimise binary decision. 

We start from a single qubit initially prepared in a pure
state.
In this case, as it was observed in Section~\ref{sec2}, the optimal NP strategy may be analytically determined and the characteristic equation $p_{11}=p_{11}(\kappa, p_{10})$ takes the form of the expression shown in Eq.~(\ref{eq:NP pure}), where $\kappa = \braket{\psi_0}{\psi_\lambda} = \langle\psi_0| U_{\lambda}|\psi_{0}\rangle$ is the overlap between the initial and the perturbed states.
Given the expression of $U(\lambda)=e^{-i \lambda\sigma_1}$, it is clear that the eigenstates of $\sigma_1$ are not modified by the action of the perturbation (up to an irrelevant phase), and thus they are not suitable to detect any value of $\lambda$.
\par
\begin{figure}[h!]
\includegraphics[width=0.45\columnwidth]{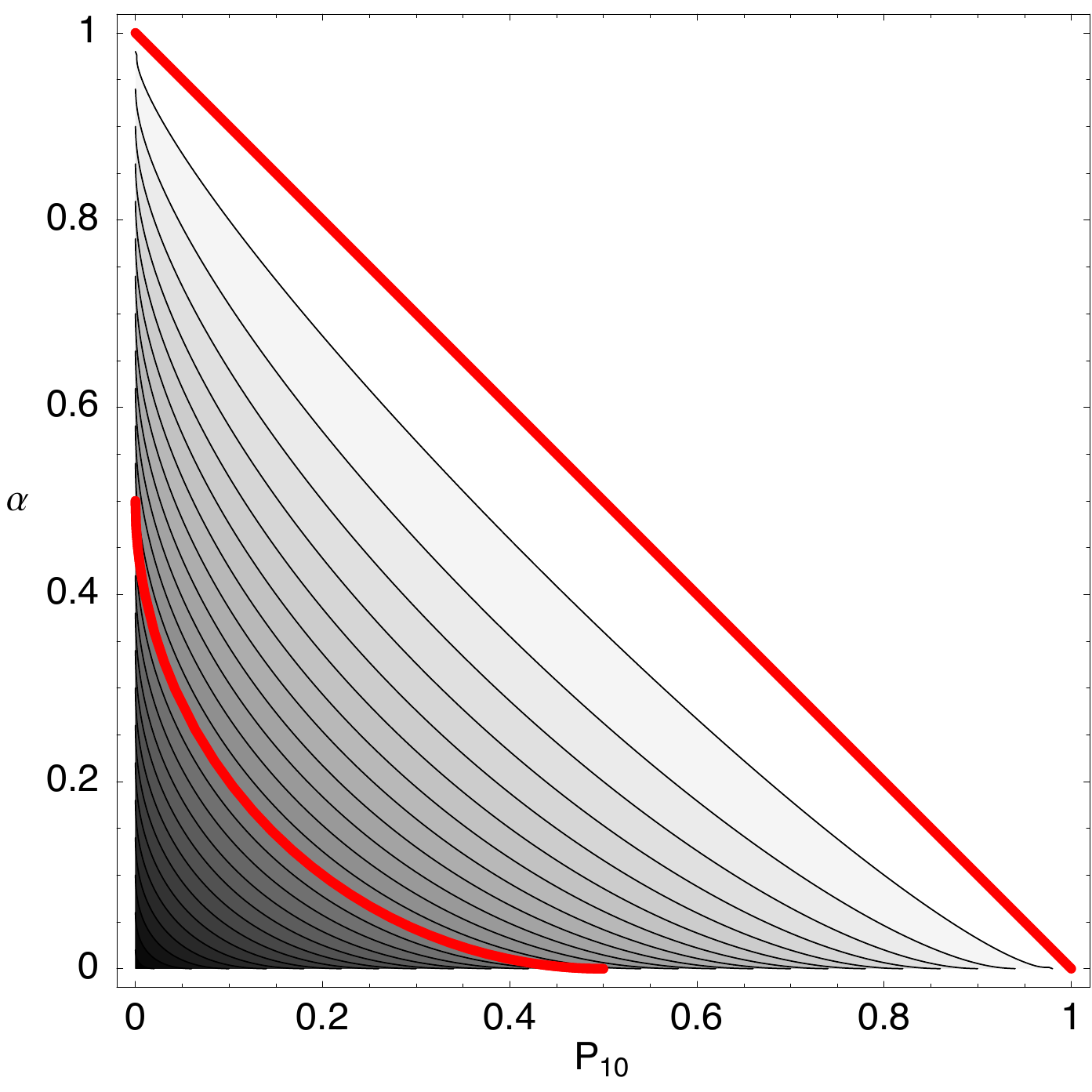}  \includegraphics[width=0.45\columnwidth]{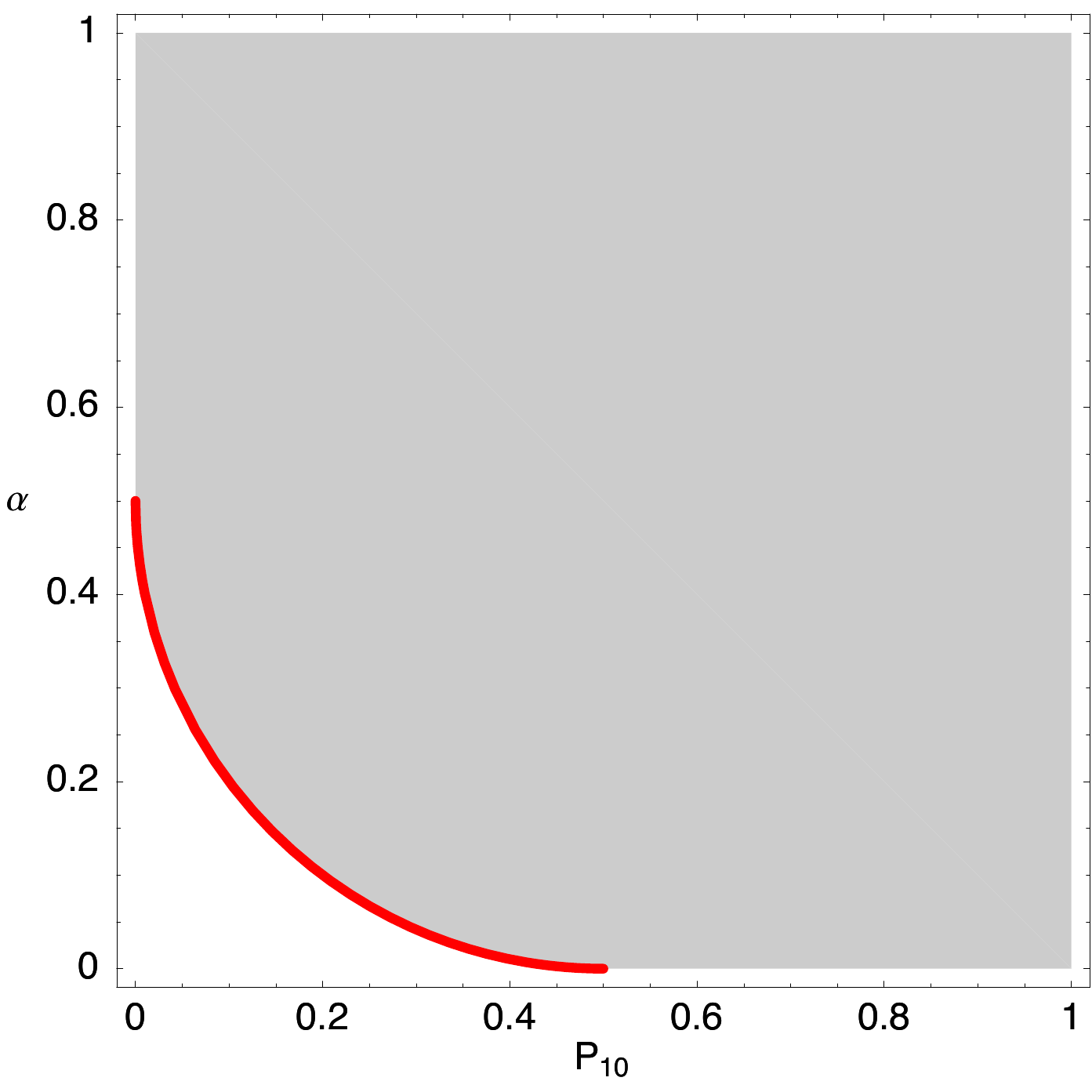}
\caption{The minimum detectable perturbation for pure states 
(absolute criterion). Left panel: contour plot $p_{11}$ 
as a function of $p_{10}$  and $\alpha$. Right: the gray 
area corresponds to the region where $p_{11} \geq \frac12$.}
\label{fig:pureA}
\end{figure}

In order to define the minimum detectable perturbation, it is necessary to define a criterion to discern the regimes for which the detection probability can be considered large.
The first criterion to be employed is an ``absolute'' one: 
a perturbation amplitude $\lambda$ is detectable if it 
leads to a detection probability $p_{11}(\kappa, p_{10})
\geq \frac12$.
A lower detection rate would indeed make the dataset 
useless, as no reliable information may be extracted 
in that case. We also address a ``relative'' criterion,
for which  a perturbation is considered detectable when 
it leads to $p_{11}/p_{10} \ge \delta \gg 1$.
In order to perform our analysis, we rewrite the expression of
$p_{11}=p_{11}(\kappa, p_{10})$ in terms of $\alpha \equiv 1- 
\left|\kappa\right|^2=\left( 1- r_1^2 \right) \sin^2\lambda$.
Concerning the expression of the detection probability, it is clear from Eq.~(\ref{eq:NP pure}) that unit detection 
probability $p_{11}=1$ is reached whenever $\left|\kappa\right|^{2} \leq p_{10}$,  i.e. $(1-r_1^2)\sin^2 \lambda 
\geq 1-p_{10}$.
In principle, in this regime $p_{11}=1$ is obtained and thus we may detect arbitrarily small perturbation.
However, the condition $\left|\kappa\right|^{2} \leq p_{10}$ is $\lambda$-dependent and this constraint poses a lower bound 
on the detectable amplitude.
This is illustrated in the left panel of {\bf Fig.~\ref{fig:pureA}}, where a contour plot of $p_{11}$ as function of $p_{10}$ and $\alpha$ is shown. 
The {\it good} area is the white one above the red straight line $\alpha = 1- p_{10}$.
\par
The other branch of the function is less trivial to analyze.
Nevertheless, it is easy to observe that $p_{11}$ increases with $p_{10}$ and $\alpha$, and then the $p_{11} \geq \frac12$ condition is satisfied in the area between the above mentioned 
line and the curve of equation $\left[ \sqrt{ p_{10} \left( 1-\alpha \right)} + \sqrt{ \left( 1- p_{10} \right) \alpha} \right]^2 = 1/2, \quad p_{10} \in \left[ 0 , \frac{1}{2} \right]$, which is the red arc of {\bf Fig.~\ref{fig:pureA}} (both panels).
The entire set of couples $\left(p_{10} , \alpha \right)$ satisfying the absolute criterion can therefore be graphically represented as the gray-shadowed area in the right panel of {\bf Fig.~\ref{fig:pureA}}.
The above Equation may be inverted to make $\alpha$ explicit, leading to $\alpha = \frac{1}{2} - \sqrt{p_{10} 
\left( 1- p_{10} \right)}, \quad  p_{10} \in \left[ 0 , 
\frac{1}{2} \right]$.
The minimum detectable perturbation can then be calculated upon recalling the expression of $\alpha$, leading to
\begin{equation}
\lambda_m = \arcsin \sqrt{\frac{1}{1-r_1^2} \left[ \frac{1}{2} - \sqrt{p_{10} \left( 1- p_{10} \right)} \right]}\,.
  \label{eq:sin2-pureA}
\end{equation}
In order to make sense of the right-hand side of Eq.~(\ref{eq:sin2-pureA}), the possible values of $r_1$ 
must be restricted to $r_1^2 \le \frac{1}{2} + \sqrt{p_{10} \left( 1- p_{10} \right)}$.
This is a limitation on the state construction: if we fix a 
false-alarm probability $p_{10}\leq\frac12$, only states satisfying this 
condition may lead to a detection probability $p_{11}$ larger than $\frac12$.
In the rest of our discussion we focus on those states, since for $p_{10} \geq \frac12 $ we have $p_{11}=1$, and any state preparations corresponds to an arbitrarily small detectable perturbation.
As it was noted above, the optimal preparation of the input signal corresponds to $r_1=0$.
In this case, varying $p_{10} \in [0, \frac12]$, we have 
$\sin^2\lambda \in [\frac12,0]$.
Accepting as instance no false alarm ($p_{10}=0$), the minimum detectable perturbation parameter is calculated as 
$\lambda_{m} = \pi/4$, whereas all the unitary 
transformations with parameter smaller than $\pi/4$ cannot be detected.    
When instead the accepted false-alarm probability is fixed to 
$1/2$ (the largest possible value when considering this branch), 
all the possible transformation parameters are available: 
$\lambda_{m} = 0$.    
This is yet a relative advantage, because of the trade-off between 
sensitivity and false-alarm probability. In particular, for values 
of $p_{10}\simeq \frac12$  we have 
$\lambda_m \simeq \sqrt{\frac{1}{1-r_1^2} \left[ \frac{1}{2} - 
 \sqrt{p_{10} \left( 1- p_{10} \right)} \right]}$.
 If $r_1$ takes its maximum allowed value, that is $r_1^2 = \frac{1}{2} + \sqrt{p_{10} \left( 1- p_{10} \right)}$, then  
 $\lambda_m=\pi/2$ independently from the value of the false-alarm probability.
\par
\begin{figure}[h!]
\includegraphics[width=0.45\columnwidth]{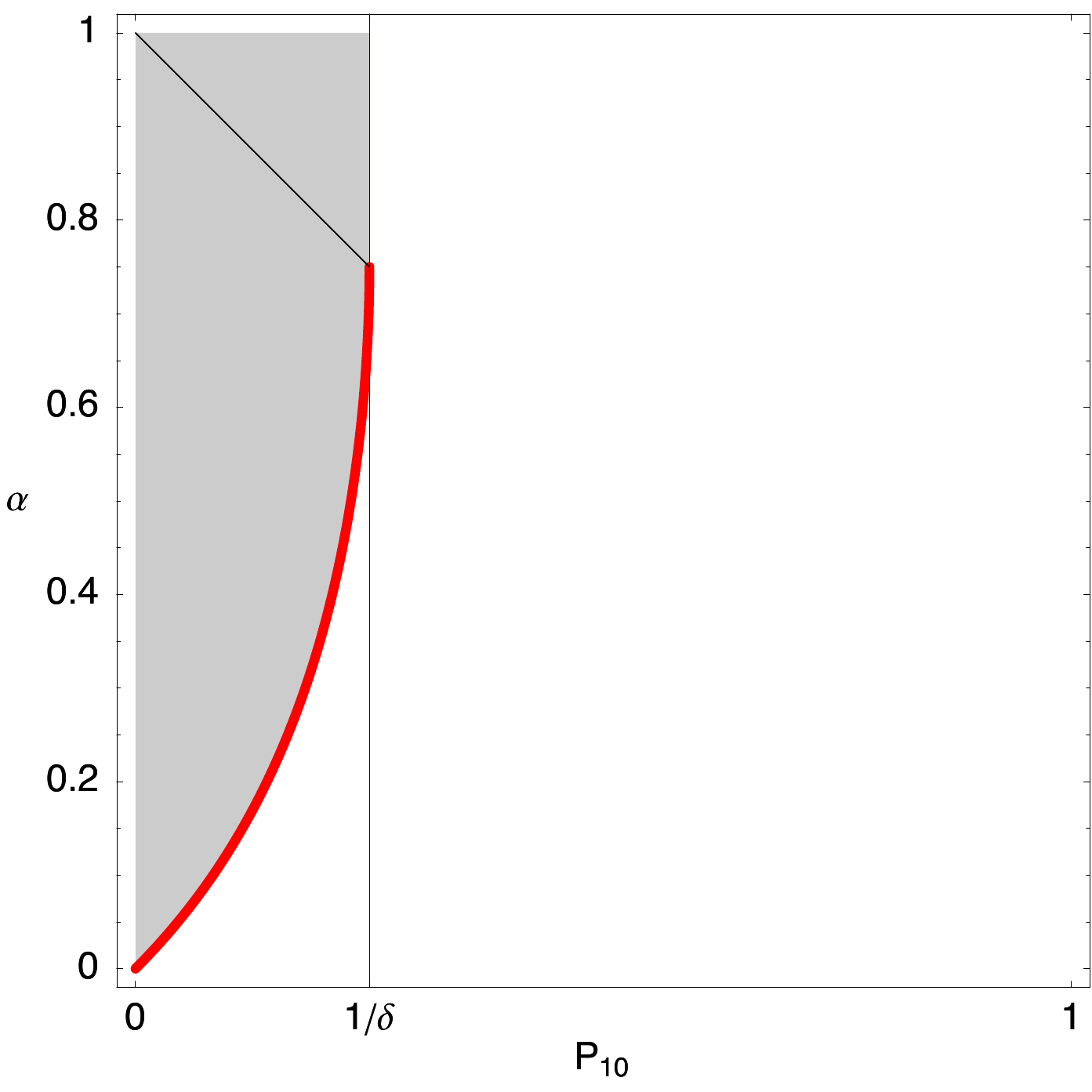}
\caption{\textcolor{black}{The minimum detectable perturbation 
for pure states (relative criterion). The gray area
corresponds to the regione where $p_{11}/p_{10}\ge \delta$ for the particular case $\delta=4$.}}
\label{fig:pureB}
\end{figure}

Let us now address the analysis of the second above-mentioned criterion, for which the minimum detectable perturbation leading to $p_{11}/p_{10} \ge \delta \gg 1$ is targeted.
As in previous cases, it is easy to address the situation 
when $\alpha \ge 1- p_{10}$, for which $p_{11} =1$.
In this case, the criterion may be easily inverted and immediately leads to $p_{10} \le \frac{1}{\delta}$.
The corresponding points 
in the $p_{10}-\alpha$ plane are shown 
in {\bf Fig.~\ref{fig:pureB}}. In the opposite 
case ($\alpha \le 1-p_{10}$), the inequality
$\left[ \sqrt{ p_{10} \left( 1-\alpha \right)} + \sqrt{ \left( 1- p_{10} \right) \alpha} \right]^2 \ge \delta$
has to be solved. After some algebra, we may rewrite this condition as
\begin{equation}
 \alpha \ge p_{10} \left[ \sqrt{1- \delta p_{10}} - \sqrt{ \delta \left(1- 
 p_{10} \right)} \right]^2\,.
\label{eq:cond_pureB}
\end{equation}
The entire area for which the detection probability can be considered large with respect to $p_{10}$ is gray coloured in {\bf Fig.~\ref{fig:pureB}}.
Upon increasing $\delta$, the area is decreasing in size, the vertical line moving left.
\par
The minimum detectable perturbation is thus given by
\begin{equation}
\lambda_m = \arcsin\sqrt{\frac{p_{10}}{1-r_1^2} \left[ \sqrt{1- \delta p_{10}} - \sqrt{ \delta \left( 1- p_{10} \right)} \right]^2},
\end{equation}
depending on the fixed value of $\delta$ and on the state preparation with respect to the $\sigma_1$-axis, $r_1$.
The restriction on the possible state preparation reads
as follows: $r_1^2 \le 1-  p_{10} \left[ \sqrt{1- \delta p_{10}} - \sqrt{ \delta \left( 1- p_{10} \right)} \right]^2$.
\subsection{Mixed states}
Following the same steps for mixed states, we are led to 
diagonalise the Lagrange characteristic operator 
$\Gamma  =\rho_{\lambda}-\gamma\rho_{0}
=\begin{pmatrix}\Gamma_{0} & \Gamma_{2}\\
\Gamma_{2}^{*} & \Gamma_{1}\end{pmatrix}$, 
which is expressed in terms of the Bloch representation 
of the initial and the perturbed states 
\begin{align}
\Gamma_{0} & =\frac12 \left[\left(1-\gamma\right)+\left(r_{3\lambda}-\gamma r_{3}\right)\right],\\
\Gamma_{1} & =\frac12 \left[\left(1-\gamma\right)-\left(r_{3\lambda}-\gamma r_{3}\right)\right],\\
\Gamma_{2} & =\frac12 \left[\left(r_{1\lambda}-\gamma r_{1}\right)-i\left(r_{2\lambda}
-\gamma r_{2}\right)\right]\,.
\end{align}
The detection and the false alarm probabilities are respectively given by:
\begin{align}
p_{11}  =\frac{1}{2}\left[1+\frac{\left(f(\gamma)-\gamma
|\kappa|^2 \right)\,\sqrt{r^2}}{\sqrt{f^{2}(\gamma)-\gamma|\kappa|^2}}\right],\label{eq:eq1} \quad
p_{10}  =\frac{1}{2}\left[1-\frac{\left(f(\gamma)-|\kappa|^2\right)\,\sqrt{r^2}}{\sqrt{f^{2}(\gamma)-\gamma|\kappa|^2}}\right],
\end{align}
where $f(\gamma) =\frac12 (1+\gamma)$ and
$|\kappa|^2  =1-\left(1-\frac{r_{1}^{2}}{r^{2}}\right)\sin^2\lambda$.
After careful inspection of the domains of validity, we may invert 
the above Equations, obtaining the characteristic expression $p_{11}=p_{11}(\kappa,p_{10})$ as follows:
\begin{equation}
p_{11}\left(p_{10}\right)=
\left\{
\begin{array}{cl}
0 & 0\leq p_{10}\left(\gamma\right)\leq p_{10}\left(\gamma_{+}\right)\\
& \\
p_{11}^*
& p_{10}\left(\gamma_{+}\right)<p_{10}<p_{10}\left(\gamma_{-}\right) \\
& \\
p_{11}\left(\gamma_{-}\right) & p_{10}\left(\gamma_{-}\right)\leq p_{10}\leq1 \\
& \\
1 & p_{10}=1 \\
& \\
\end{array}\right.\,, \label{eq:charact}
\end{equation}
where 
$p_{11}^*=p_{10}|\kappa|^2+\left(1-p_{10}\right)\left(1-|\kappa|^2\right)
+\sqrt{|\kappa|^2\left(1-|\kappa|^2\right)\left[r^{2}-\left(2p_{10}-1\right)^{2}\right]} $, 
and the two critical values $\gamma_\pm$ are given by
\begin{equation}
\gamma_\pm = 1+ 2 \frac{\sqrt{r^2(1-|\kappa|^2)}}{1-r^2} \left[ \sqrt{r^2 -|\kappa|^2 r^2} \pm \sqrt{1-|\kappa|^2 r^2} \right]\,.
\end{equation}
\begin{figure}[h!]
\includegraphics[width=0.75\columnwidth]{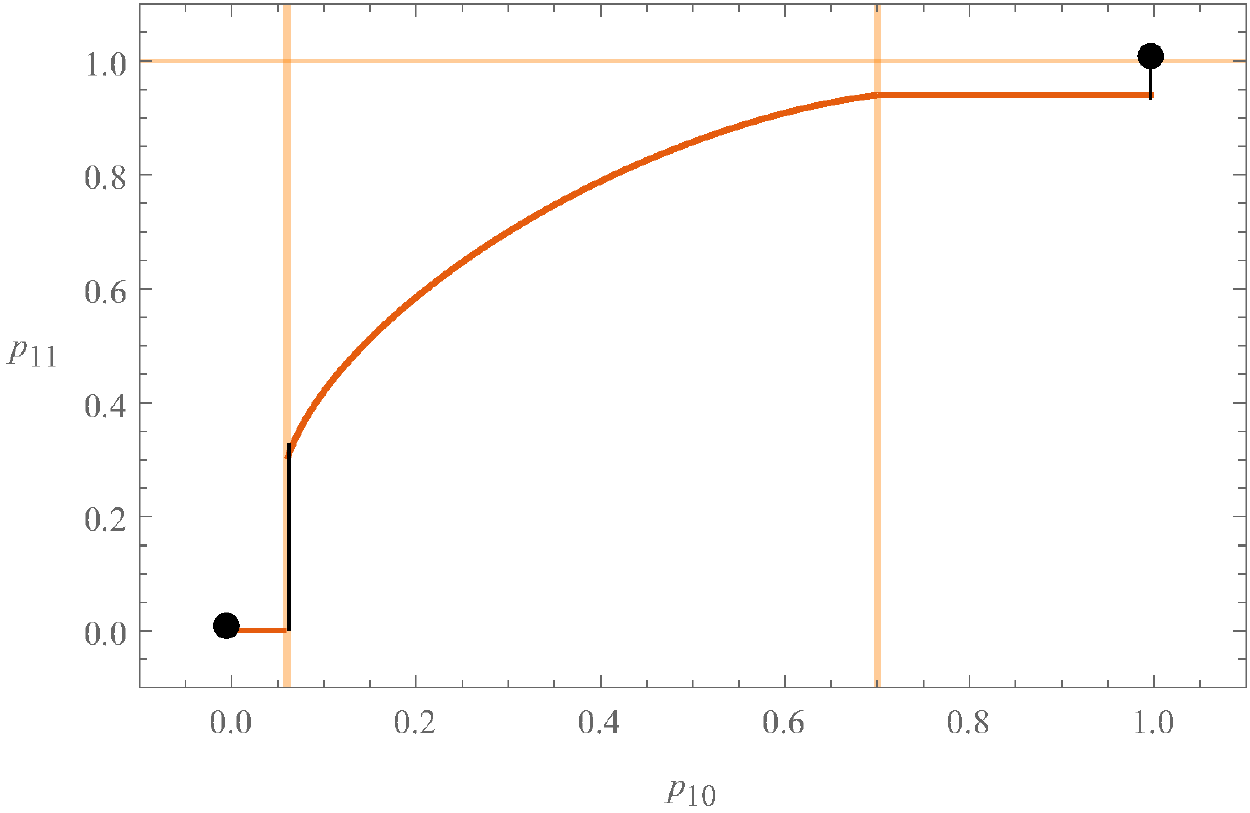} 
\caption{The minimum detectable perturbation for mixed states. 
The plot shows the characteristic function of Eq. (\ref{eq:charact}) 
\textcolor{black}{for an initially} mixed state with $r^{2}$=$0.8$
and $|\kappa|^2$=$0.8$. \label{fig:The-characteristic-fucntion}}
\end{figure}
\par
The characteristic function is graphically represented in \textbf{Fig.~\ref{fig:The-characteristic-fucntion}} 
for fixed values of $r^{2}$ and $\kappa$, which correspond to fix $r_{1}^{2}$ and $\lambda$, i.e. to tuning the initial preparation and the setup (remind that $r^2= 2\mu-1$, where $\mu$ is the purity of the initial preparation).
The corresponding minimum detectable perturbation is then given by 
\begin{equation}
\lambda_m=\arcsin \sqrt{\frac{1}{1-\frac{r_{1}^{2}}{r^{2}}}\left[\frac{1}{2}-\sqrt{\frac{r^{2}-1+4p_{10}\left(1-p_{10}\right)}{r^{2}}}\right]}\,,\label{eq:sin}
\end{equation}
with the condition 
\begin{equation}
r_{1}^{2}\leq r^{2}\left[\frac{1}{2}+\sqrt{\frac{r^{2}-1+4p_{10}\left(1-p_{10}\right)}{r^{2}}}\right]\,.\label{eq:r max}
\end{equation}
Once again, this is a restriction on the initial preparation of the
system. When the inequality is saturated, the only parameter that
can be detected is $\lambda_{m}=\frac{\pi}{2}$, otherwise the best
preparation is $r_{1}=0$. Of course, $\lambda_{m}$ vanishes as far as 
$p_{10}$ approaches the limiting value $p_{10}=\frac{1}{2}$.
\subsection{Pure two-qubit states}
Here we address possible enhancement coming from coupling 
the qubit that might undergo a perturbation $U_{\lambda}$
to another qubit which is left unperturbed. We will consider 
the same setup of the Bayesian analysis, reported in 
in \textbf{Fig.~\ref{fig:The-experimental-protocol}}.
We obtained before that for a generic pure state described 
in the computational basis of two qubits by (\ref{eq:general 2-qubit}) 
and satisfying $\alpha_0\alpha_2^{*}+\alpha_1\alpha_{3}^{*}=0$ the 
overlap has a universal 
value independent of the preparation; $\left|\kappa\right|^{2}=\cos^2\lambda$,
and thus $\alpha=\sin^{2}\lambda$.
If we compare this result with what we found in the case of generic
single qubit state, we see that they match up whenever the state
is pure and the projection on the generator axis vanishes.
The corresponding minimum detectable perturbation is given by
\begin{equation}
\lambda_m=\arcsin \sqrt{\frac{1}{2}-\sqrt{p_{10}\left(1-p_{10}\right)}}
\label{eq:not}\,,
\end{equation}
which does not depend on the initial preparation of the system
and ranges from $0$ to $\frac{\pi}{4}$ for
$p_{10}$ ranging from $\frac{1}{2}$ to $0$. As we already concluded for the Bayesian case, the effect of entanglement is to enhance the overall stability of the discrimination/estimation scheme.

\subsection{Mixed two-qubit states}
\begin{figure}[h!]
\includegraphics[width=0.7\columnwidth]{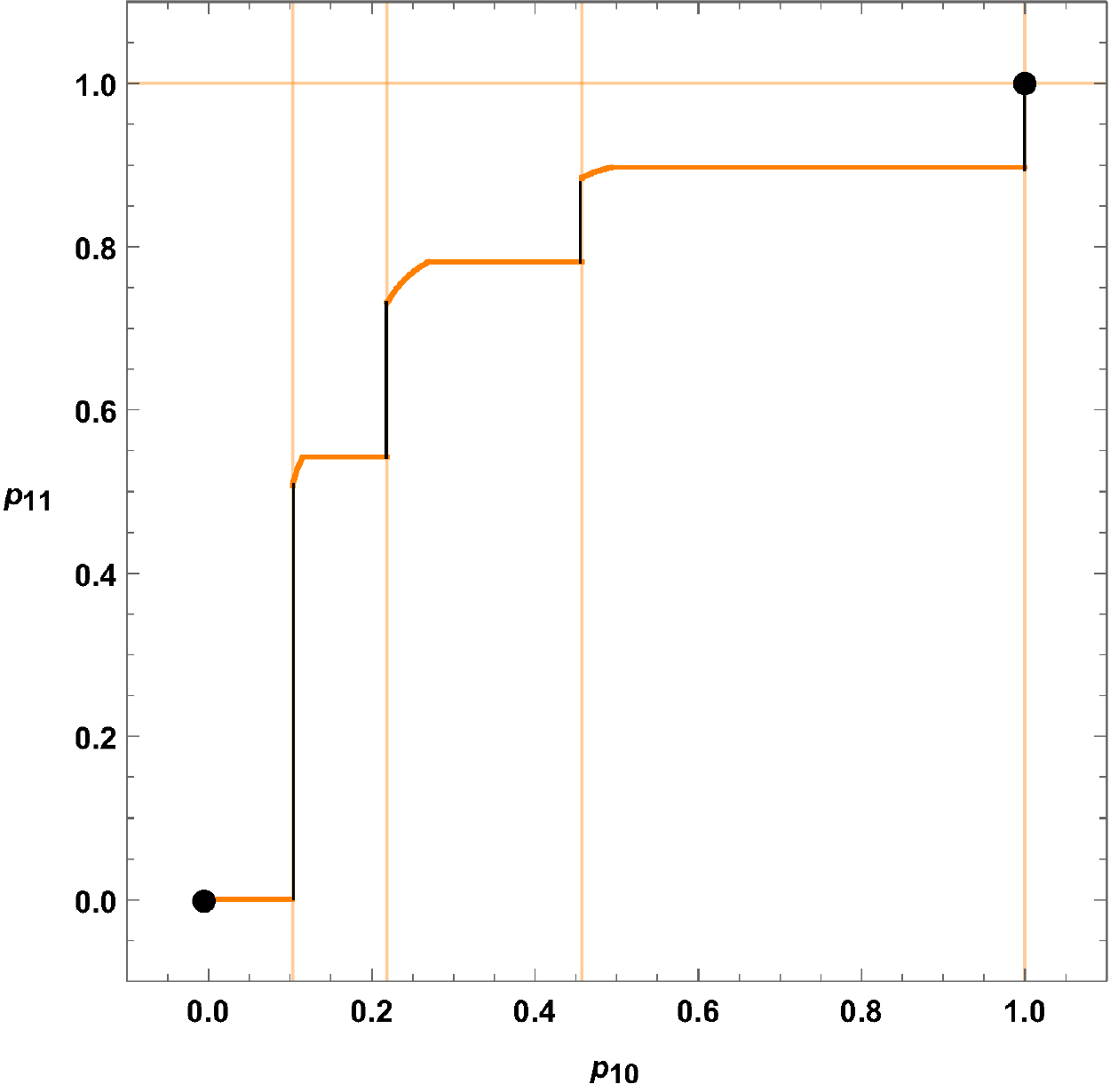} 
\caption{\textcolor{black}{The characteristic function of Eq. (\ref{eq:char prob})}
for a diagonal mixed state with coefficients $p_{0}=0.1$, 
$p_{1}=0.2$, $p_{2}=0.1$ and $p_{3}=0.6$.
\label{fig:The-characteristic-function-1}}
\end{figure}

Let us consider here a generic Bell-diagonal mixed state \textcolor{black}{$\rho_{0}=p_{0}\left|\phi_{+}\rangle\langle\phi_{+}\right|+p_{1}\left|\psi_{+}\rangle\langle\psi_{+}\right|+p_{2}\left|\psi_{-}\rangle\langle\psi_{-}\right|\nonumber +p_{3}\left|\phi_{-}\rangle\langle\phi_{-}\right|$, $\sum_{k}p_{k}=1$.} 
The Lagrange operator $\Gamma=\rho_\lambda -\gamma \rho_0$ for this preparation 
may be decomposed into a direct sum of two operators, each one 
acting on the orthogonal subspaces generated by $\left\{|\phi_{+}\rangle, |\psi_{+}\rangle\right\}$
and $\left\{|\phi_{-}\rangle, |\psi_{-}\rangle\right\} $ respectively.    
In particular, we have
\begin{equation}
\mathbf{\Gamma}=\mathbf{\Gamma}^{\left(+\right)}\oplus\mathbf{\Gamma}^{\left(-\right)},
\end{equation}
where
\textcolor{black}{\begin{align}
	\mathbf{\Gamma}^{\left(+\right)}= & \left[p_{0}\left(\cos^{2}\lambda-\gamma\right)+p_{1}\sin^{2}\lambda\right]\left|\phi_{+}\rangle\langle\phi_{+}\right|\nonumber \\ & +\left[p_{1}\left(\cos^{2}\lambda-\gamma\right)+p_{0}\sin^{2}\lambda
	\right]\left|\psi_{+}\rangle\langle\psi_{+}\right|\nonumber \\
	& +i\,\sin\lambda \cos\lambda\left(p_{0}-p_{1}\right)\left(\left|\phi_{+}\rangle\langle\psi_{+}\right|-\left|\psi_{+}\rangle\langle\phi_{+}\right|\right),
	\end{align}} 
and
\textcolor{black}{\begin{align}
	\mathbf{\Gamma}^{\left(-\right)}= & \left[p_{2}\left(\cos^{2}\lambda-\gamma\right)+p_{3}\sin^{2}\lambda\right]\left|\phi_{-}\rangle\langle\phi_{-}\right|\nonumber \\ & +\left[p_{3}\left(\cos^{2}\lambda-\gamma\right)+p_{2}\sin^{2}\lambda\right]\left|\psi_{-}\rangle\langle\psi_{-}\right|\nonumber \\
	&+i\,\sin\lambda \cos\lambda\left(p_{2}-p_{3}\right)\left(\left|\phi_{-}\rangle\langle\psi_{-}\right|-\left|\psi_{-}\rangle\langle\phi_{-}\right|\right).
	\end{align}} 
The two operators may be brought to a $2\times 2$ matrix form 
as
\begin{equation}
\mathbf{\Gamma}^{(\pm)} = \left( \begin{array}{cc} \Gamma_0^{(\pm)} & \Gamma_2^{(\pm)} \\ \left( \Gamma_2^{(\pm)} \right)^* & \Gamma_1^{(\pm)} \end{array} \right),
\end{equation}  with 
\begin{align}
\Gamma_{0}^{\left(+\right)} & =p_{0}\left(\cos^{2}\lambda-\gamma\right)+p_{1}\sin^{2}\lambda, &\quad \Gamma_{0}^{\left(-\right)} & =p_{2}\left(\cos^{2}\lambda-\gamma\right)+p_{3}\sin^{2}\lambda, \nonumber \\
\Gamma_{1}^{\left(+\right)} & =p_{1}\left(\cos^{2}\lambda-\gamma\right)+p_{0}\sin^{2}\lambda, & \quad \Gamma_{1}^{\left(-\right)} & =p_{3}\left(\cos^{2}\lambda-\gamma\right)+p_{2}\sin^{2}\lambda,
\nonumber \\
\Gamma_{2}^{\left(+\right)} & =i\,\sin\lambda \cos\lambda \left(p_{0}-p_{1}\right),
&\quad
\Gamma_{2}^{\left(-\right)} & =i\sin\lambda \cos\lambda\left(p_{2}-p_{3}\right)\,.
\end{align}
For the sake of clarity, let us discuss explicitly 
how to derive the characteristic function. 
The relevant quantities for discussing the eigenvalues of $\mathbf{\Gamma}$ are trace and determinant of $\mathbf{\Gamma}^{\left(\pm\right)}$.  We have
\begin{align}
\hbox{Tr}\left[\mathbf{\Gamma}^{\left(+\right)}\right] & =\left(p_{0}+p_{1}\right)\left(1-\gamma\right) \\ 
\det\mathbf{\Gamma}^{\left(+\right)}  & = p_{0}p_{1}\left[\gamma^{2}-2\gamma\left(1+\Xi^{\left(+\right)}\right)+1\right],
\end{align}
where
\begin{equation}
\Xi^{\left(+\right)}=\frac{\left(p_{0}-p_{1}\right)^{2}}{2p_{0}p_{1}}\mathrm{sin^{2}}\lambda\,.
\end{equation}
The two critical values of $\gamma$, corresponding to vanishing determinant, are 
\begin{equation}
\gamma_{\pm}^{\left(+\right)}=1+\Xi^{\left(+\right)}\pm\sqrt{\Xi^{\left(+\right)}\left(\Xi^{\left(+\right)}+2\right)}\,.
\end{equation}
Accordingly, three regimes are identified, summarized as follows:
\textcolor{black}{\begin{align}
	\left\{
	\begin{array}{lcl}
	\gamma^{(+)}\leq\gamma_{-}^{(+)}, \zeta_{\pm}\geq 0 &\quad& \Pi=\left\{ \Pi_{0}=\mathbb{O}\,,\Pi_{1}=\mathbb{I}_4\right\} \\
	\gamma^{(+)}\geq\gamma_{+}^{(+)} , \zeta_{\pm}\leq 0
	&\quad& \Pi=\left\{ \Pi_{0}=\mathbb{I}_4\,,\Pi_{1}=\mathbb{O}\right\} \\
	\gamma_{-}^{(+)}<\gamma^{(+)}<\gamma_{+}^{(+)}, \zeta_{-}\leq0\,;\,\zeta_{+}>0
	&\quad& \Pi=\Bigl\{\Pi_{0}=\left|\zeta_{-}\rangle\langle\zeta_{-}\right|,
	\Pi_{1}=\left|\zeta_{+}\rangle\langle\zeta_{+}\right|\Bigr\}
	\end{array}\right.
	\end{align}
where $\zeta_{\pm}$ are the eigenvalues of the operator $\mathbf{\Gamma}^{\left(+\right)}$.} 
Upon carrying out the same analysis for $\mathbf{\Gamma}^{\left(-\right)}$, we identify the critical values for the Lagrange multiplier.
Depending on the initial parameters $\left\{ p_{i}\right\} _{i=0,1,2,3}$, the intervals $\left[\gamma_{-}^{\left(-\right)},\gamma_{+}^{\left(-\right)}\right]$
and $\left[\gamma_{-}^{\left(+\right)},\gamma_{+}^{\left(+\right)}\right]$ are contained one in the other.
Focusing on the case $\Xi^{\left(+\right)}\geq\Xi^{\left(-\right)}$, it is observed that $\left[\gamma_{-}^{\left(-\right)},\gamma_{+}^{\left(-\right)}\right]\subseteq\left[\gamma_{-}^{\left(+\right)},\gamma_{+}^{\left(+\right)}\right]$.
The characteristic function may then be explicitly parameterized by $\gamma$ as follows:  
\begin{equation}
\begin{cases}
p_{10}=1 & p_{11}=1\\
\\
p_{10}=\frac{1}{2}\bigg[\left(p_{0}+p_{1}\right)+\left|p_{0}-p_{1}\right|\times & p_{11}=\frac{1}{2}\bigg[\left(p_{0}+p_{1}\right)+\left|p_{0}-p_{1}\right|\times\\
\qquad \frac{\cos2\lambda-\gamma}{\sqrt{\gamma^{2}-2\gamma\cos2\lambda+1}}\bigg] +\left(p_{2}+p_{3}\right) & \qquad \frac{1-\gamma\cos2\lambda}{\sqrt{\gamma^{2}-2\gamma\cos2\lambda+1}}\bigg]+\left(p_{2}+p_{3}\right)\\
\\
p_{10}=\frac{1}{2}\bigg[1+\left(\left|p_{2}-p_{3}\right|+\left|p_{0}-p_{1}\right|\right)\times & p_{11}=\frac{1}{2}\bigg[1+\left(\left|p_{2}-p_{3}\right|+\left|p_{0}-p_{1}\right|\right)\times\\
\qquad \frac{\cos2\lambda-\gamma}{\sqrt{\gamma^{2}-2\gamma\cos2\lambda+1}}\bigg] & \qquad \frac{1-\gamma\mathrm{cos2\lambda}}{\sqrt{\gamma^{2}-2\gamma\cos2\lambda+1}}\bigg] 
\\ \\
p_{10}=\frac{1}{2}\bigg[\left(p_{0}+p_{1}\right)+\left|p_{0}-p_{1}\right|\times & p_{11}=\frac{1}{2}\bigg[\left(p_{0}+p_{1}\right)+\left|p_{0}-p_{1}\right|\times\\
\qquad \frac{\mathrm{cos2\lambda}-\gamma}{\sqrt{\gamma^{2}-2\gamma\mathrm{cos2\lambda}+1}}\bigg] & \qquad \frac{1-\gamma\mathrm{cos2\lambda}}{\sqrt{\gamma^{2}-2\gamma\mathrm{cos2\lambda}+1}}\bigg]
\\ \\
p_{10}=0 & p_{11}=0
\end{cases}\label{eq:char prob}
\end{equation}
An example of this characteristic function with a particular
choice of the preparation parameters $p_{0},p_{1},p_{2},p_{3}$ and the perturbation parameter $\lambda$ is illustrated in 
\textbf{Fig.~\ref{fig:The-characteristic-function-1}}.
Similar calculations can be conducted in the other case for which: $\Xi^{\left(-\right)}\geq\Xi^{\left(+\right)}$.
\par
Using arguments similar to those developed in this Section we may deal with discrimination in the presence of noise and find the minimum detectable perturbation also in those cases. Finally, we notice that the use of entanglement can positively affect the discrimination process of quantum states by confirming the overall precision obtained in the single-qubit protocol, while enhancing the stability of the detection scheme by removing the dependence on the preparation of the initial state.
\section{Conclusions}\label{outro}
In this paper, we have applied concepts and methods from 
quantum decision theory to address process discrimination 
and estimation in qubit systems. In particular, we have discussed 
the problem of discriminating whether or not a given 
unitary perturbation was applied to a qubit system, 
as well as the complementary problem of evaluating the 
minimum detectable perturbation amplitude leading to discriminable
outputs. Our approach has allowed us to obtain several results 
which are summarised in the following.
\par
We have seen that entanglement may represent a resource, 
in particular for improving the stability of the discrimination 
strategies. Using single-qubit pure probes, the characteristic 
function for both Bayes and NP strategy does explicitly depend 
on the initial preparation, whereas using two-qubit probes
this dependence disappears. Bell-state probes are shown to be 
optimal for both strategies. Nevertheless, they are not the only 
choice, since there exists an entire class of states, depending 
on the generator of the unitary perturbation, for which both 
strategies are optimised. In this sense, it has been shown that 
not only Bell states belong to this class, but there exists 
also a set of non-balanced states simultanously optimal for 
each generator.
\par
A similar analysis has been performed for single- and two-qubit 
mixed states, for which we found novel analytic solutions for 
Bayes error probability and the NP characteristic function. In 
this way, we found that optimal preparations correspond to pure 
states with a vanishing value of the generator expectation 
value. We have then used our results to explicitly discuss the 
effects of possible background noises on the discrimination 
performance of the various strategies and schemes.
Overall, besides the exact quantification of noise effects, 
we have the rather intuitive conclusion that when we bring noise 
into play, it has detrimental effects only when it acts on the 
eigenspaces of the generator, whereas it has no effects when acting 
on orthogonal subspaces.
\section*{Acknowledgements}
MGAP is member of GNFM-INdAM.

\end{document}